\UseRawInputEncoding
\documentclass[12pt, a4paper]{article}
\title{The Barbero - Immirzi Parameter: An Enigmatic Parameter of Loop Quantum Gravity}
\author{Rakshit P. Vyas$ ^1 $, Mihir J. Joshi$ ^2 $}
\date {%
$^1$ \textit{Department of Physics, Saurashtra University, Rajkot, India} \\%
$^2$ \textit{Department of Physics, Saurashtra University,  Rajkot, India} \\%
\today %
}

\usepackage{blindtext}
\usepackage{geometry}
\usepackage{amsmath}
\usepackage{amssymb}
\usepackage{amsfonts}
\usepackage{graphicx}
\usepackage{stabular}
\begin{document}

\maketitle

\begin{abstract}

The Barbero - Immirzi parameter (\(\gamma\)) is introduced in loop quantum gravity (LQG) whose physical significance is still a biggest open question; because of its profound traits. In some cases, it is real-valued; while, it is complex-valued in other cases. This parameter emerges out in the process of denoting a Lorentz connection with non compact group \(SO(3,1)\) in the form of a complex connection with values in a compact group of rotations, either \(SO(3)\) or \(SU(2)\). Initially, it was appeared in the Ashtekar variables.  Fernando Barbero proposed its possibility to include within formalism.  Its present value is fixed by counting of micro states in loop quantum gravity and matching with the semi classical black hole entropy computed by Stephen Hawking. This parameter is used to count the size of the quantum of area in Planck units.  Until, the discovery of the spectrum of the area operator in LQG; its significance remains unknown. However, its complete physical significance is yet to be explored. In the present article, an introduction to the Barbero - Immirzi parameter in LQG, time line of this research area, various proposals regarding its physical significance are given. 

Keywords: Loop quantum gravity, Ashtekar variable, Barbero-Immirzi parameter, area operator, black hole entropy 

\end{abstract}

\tableofcontents

\section{Introduction}

Loop quantum gravity (LQG) is one of the supposed candidate of theory of quantum gravity. It can unify general relativity (GR) with quantum field theory (QFT). 
It is non - perturbative and background independent approach of quantum gravity theory. LQG begins with GR; thereafter, takes some conceptual base from QFTs to deliver a quantum theory of gravity. LQG is a theory of constraints, in which various constraints such as Hamiltonian, diffeomorphism and Gauss constraints are converted into operators. In the canonical quantization approach of LQG (ADM formalism), 3+1 decomposition of spacetime is necessary to quantize gravity; however, the covariant approach (sum over geometry) follows different strategy. Due to limited space, basics of LQG is not given. There are many classic texts [ 1 - 10] and papers [11 -23] that explain LQG lucidly.

In 1986, Abhay Ashtekar [24] found new kind of variables (Ashtekar’s variable) in classical and quantum gravity. In Ashtekar's formulations, the constraints are simplified by considering a complex-valued form for the connection and tetrad variables and these are known as Ashtekar's variables [1 - 5]. 

While dealing with the reality condition of the formalism of Ashtekar's variable, Barbero [25-26] firstly, introduced a free parameter in the expression of Ashtekar's variable and then in the expression of constraints. Thereafter, Immirzi [27 - 28] used various possibilities of this free parameter in the expression of LQG in his paper. This free parameter is now known as the Barbero - Immirzi parameter (\(\gamma\)) (in short, BI parameter).  Whether the BI parameter \(\gamma\) is complex or real; it gives many results in LQG. In some cases, the real-valued BI parameter \(\gamma\) is required; while, in the other cases the complex-valued BI parameter \(\gamma\) is necessary [1 - 5].

The physical significance of the area operator in LQG with the complex BI parameter \(\gamma\) becomes ambiguous. The LQG kinematics i.e., kinematical Hilbert space can only be comprehended; if, the BI parameter \(\gamma\) is real number. The \(SU(2)\) spin network of LQG can only be created with the real value of the BI parameter \(\gamma\) [1 - 5].

With the complex value of the BI parameter \(\gamma\)(\(\gamma = i\)),  the spatial connection can be seen as spacetime connection; since, it transforms under diffeomorphism in right way.  There are also some cases that shows that the complex valued BI parameter \(\gamma\) is also crucial in LQG formalism. For instance, the form of Hamiltonian constraints becomes simpler; if,  \(\gamma = i\) is taken [1 - 5].

In the next section, various proposals regarding the BI parameter are briefly explained; in which some proposals advocates the real valued BI parameter \(\gamma\), while, the other advocate complex valued BI parameter \(\gamma\).   

\subsection{Ashtekar's formalism}

Before the discovery of Ashtekar's variables, the Palatini action i.e., the first order formulation was not complete . But, Ashtekar formalism made it complete. In the Palatini action, the tetrad \(e_{J}^{\mu}\) and the spin connection \(\omega_{\mu}^{JK}\) are used as independent variables. In general relativity, The Palatini action is written as [1-5]

\begin{eqnarray}
S_{P} = \int d^{4} x e e_{J}^{\mu}e_{K}^{\eta} \Omega_{\mu\eta}^{JK}[\omega]
\end{eqnarray}

Where \(e=\sqrt{-g}\) and \(\Omega_{\mu\eta}^{JK}\) is the curvature. Using the Palatini action, the Einstein field equation could be derived; but, the form of equation of constraints within this formalism, is mathematically complicated [1-5, 12, 15, 17]. The generalization of the Palatini action is equivalent to the Ashtekar formalism and it is achieved by the Holst action [29].  

In the Ashtekar formalism, by converting tetrads into triads i.e. three dimensional hypersurfaces \(\Sigma_{t}\); one gets \(e_{\mu}^{J} \rightarrow e_{c}^{j}\) where, \(\mu \rightarrow c \in \lbrace 1, 2, 3\rbrace\), \(J \rightarrow j \in \lbrace 1, 2, 3 \rbrace\) and the spin connection is also transformed  as \(\Gamma_{c}^{j} = \omega_{ckl}\varepsilon^{klj}\) [1-5, 12, 15, 17].

The Hamiltonian constraint is a complicated non-polynomial function in Palatini formulation; thus, canonical quantization is not easy within this formalism. In Palatini formulation, the variables of phase space are \((e_{c}^{j}, \Gamma_{c}^{j})\); where, \(e_{c}^{j}\) is the intrinsic metric of the spacelike manifold \(\Sigma\) and \(\Gamma_{c}^{j}\) is a function of its extrinsic curvature [1-5, 12, 15, 17].

In Ashtekar’s formalism, complex valued connection \(\Gamma_{c}^{j}\) replaces the real connection \(\omega_{\mu}^{JK}\) with duality (either self  (+1) or anti-self (-1)) [1-5, 12, 15, 17].

\begin{eqnarray}
\tilde{E}_{j}^{c} \rightarrow \frac{1}{i} \tilde{E}_{j}^{c}, K_{c}^{j} \rightarrow A_{c}^{j} = \Gamma_{c}^{j} - i K_{c}^{j} 
\end{eqnarray}

Where, \(\tilde{E}_{j}^{c}\) is the scalar density or triad electric field, \(A_{c}^{j}\) is the Ashtekar-Barbero connection or spatial connection, \(K_{c}^{j} = k_{cd}e^{dj}\) with \(k_{cd}\) the extrinsic curvature of \(\Sigma\). Thus, there are two phase space variables i.e. \(A_{c}^{j}\) and \(\tilde{E}_{j}^{c}\) [1-5, 12, 15, 17].

Since, the Ashtekar’s connection formulation variables, i.e. \(A_{c}^{j}\) and \(\tilde{E}_{j}^{c}\)   follows rotation of \(SU(2)\) symmetry with respect to the internal indices; the Ashtekar’s formalism plays the role of \(SU(2)\) gauge theory and this \(SU(2)\) group is a subgroup of \(SL(2,\mathbb{C})\) [1-5, 12, 15, 17].

All three constraints are simplified in Ashtekar’s variables and their expression are [1-5, 12, 15, 17]

\begin{eqnarray}
\mathcal{G}_{j} = D_{c} \tilde{E}_{j}^{c}
\end{eqnarray}

\begin{eqnarray}
\mathcal{C}_{c} = \tilde{E}_{j}^{d} F_{cd}^{j} - A_{c}^{j}\mathcal{G}_{j}
\end{eqnarray}

\begin{eqnarray}
\mathcal{H} = \varepsilon_{l}^{jk} \tilde{E}_{j}^{c} \tilde{E}_{k}^{d} F_{cd}^{l}
\end{eqnarray}

The equations (3), (4), and (5) are Gauss, diffeomorphism, and Hamiltonian constraint respectively. In Ashtekar's formalism, the Einstein-Hilbert-Ashtekar Hamiltonian of GR is written as [1-5, 12, 15, 17].

\begin{eqnarray}
\mathcal{H}_{EHA} = N^{c}\mathcal{C}_{c} + N \mathcal{H} + T^{j}\mathcal{G}_{j} = 0
\end{eqnarray}   

Where, \( \mathcal{C}_{c}, \mathcal{H},  \mathcal{G}_{j}, N^{c}\) and \(N \) are vector constraint, the scalar constraint,  and Gauss constraints, shift and lapse,  respectively. The \(T^{j}\) is  a Lie algebra valued function over spatial surface [1-5, 12, 15, 17]. 

The unit imaginary i.e. \(i = \sqrt{-1}\) appeared in the equation (2)  makes the formalism complex valued. Therefore, some restrictions in terms of reality condition on the possible solutions of the theory must be applied to achieve tangible  physical results relevant to a metric valued in \(\mathbb{R}\) instead of in \(\mathbb{C}\) [1-5, 12, 15, 17]. 

For example, If \(\dot{Z}\) is used to represent the time derivative of \(Z\); then, the reality condition and constraints i.e., \(\mathcal{G}_{j} = D_{c} \tilde{E}_{j}^{c}\)  must be satisfied by  solutions. In this case, there are two reality condition and the second condition is the time derivative of first condition. thus [1-5, 12, 15, 17],

\begin{eqnarray}
\tilde{E}_{j}^{c}\tilde{E}_{k}^{d}\delta^{jk} \in \mathbb{R}
\end{eqnarray}

\begin{eqnarray}
\lbrace \tilde{E}_{j}^{c}\tilde{E}_{k}^{d}\delta^{jk}\rbrace^{\bullet} \in \mathbb{R}
\end{eqnarray}

In standard form, the Ashtekar variables is given as [1-5, 12, 15, 17],

\begin{eqnarray}
\tilde{E}_{j}^{c} \rightarrow \frac{1}{\gamma} \tilde{E}_{j}^{c}, K_{c}^{j} \rightarrow A_{c}^{j} = \Gamma_{c}^{j} - \gamma K_{c}^{j}
\end{eqnarray}

Where, the \(\gamma\) is the BI parameter. If, \(\gamma = i \); then, the equation takes the original form. The Poisson brackets is written as [1-5, 12, 15, 17],

\begin{eqnarray}
\lbrace K_{c}^{j}(x), \tilde{E}_{k}^{d}(y)\rbrace = \lbrace A_{c}^{j}(x), \tilde{E}_{k}^{d}(y)\rbrace = k \delta_{k}^{j}\delta_{c}^{d}\delta(x, y)
\end{eqnarray}

In standard form [1-5, 12, 15, 17], 

\begin{eqnarray}
 \lbrace A_{c}^{j}(x), \tilde{E}_{k}^{d}(y)\rbrace = 8 \pi G \gamma \delta_{k}^{j}\delta_{c}^{d}\delta^{3}(x, y)
\end{eqnarray}

Where, \(k = 8 \pi G \gamma\).

The reality condition is not necessary for real value and as a result new variable and constraints are also real [1-5, 12, 15, 17].  

The form of Hamiltonian constraint becomes complicated with the real value of the BI parameter \(\gamma\) i.e., 

\begin{eqnarray}
\mathcal{H} = \varepsilon_{l}^{jk} \tilde{E}_{j}^{c} \tilde{E}_{k}^{d} F_{cd}^{l} - 2(1 + \gamma^{2}) \tilde{E}_{j}^{[c} \tilde{E}_{k}^{d]}K_{c}^{j}K_{d}^{k} \approx 0 
\end{eqnarray}   

If, \(\gamma = i\); then the form of Hamitonian constraints becomes simple [1-5, 12, 15, 17].

\subsection{Why the BI parameter was introduced in LQG?}

As mentioned, complex valued Ashtakar's variables simplified constraints of quantum gravity based on canonical quantization i.e., LQG. Thereafter, Barbero [25-26] came up with new strategy to tackle Ashtekar's variable with real value for Lorentzian signature space-times. In his paper, Barbero wrote down Ashtekar's variable with a free parameter (eqn. (8)); namely, \(\gamma\) (in his paper he denoted it as \(\beta\)). Ashtekar used \(SU(2)\) and \(SL(2, \mathbb{C})\)  groups of Yang - Mill theory to deliver complex valued constraints i.e., Ashtekar's variables. While, Barbero showed that one can use \(SO(3)\) Yang - Mill  phase space to write that modified Hamiltonian constraint with Lorentz signatures without complex variable to elaborate space-times without losing the features of Ashtekar's variables [1-5, 25]. 

Barbero [25-26] also explained that for simple form of Hamiltonian constraint, complex variable is required; while, for complicated form, this constraint could be written with real variables. For instance, loop variable of LQG. Barbero derived Hamiltonian constraint with \(\gamma^{2} = 1\) (real valued and Euclidean signature). While, with Lorentzian signature yields again complex valued form of equation. Barbero also derived Hamiltonian constraint with \(\gamma = - 1\). The Hamiltonian constraint could also be written with real Ashtekar variables for Lorentzian general relativity with \(SO(3)\) ADM formalism [25]. 

Thereafter, Immirzi [27 -28] explained the importance of the this parameter in his paper. In these papers, Immirzi explained canonical quantization of gravity i.e., LQG with Regge calculus briefly. In his paper, Immirzi elaborated basics of LQG briefly with the discussion on the BI parameter \(\gamma\). Immirzi discussed various possibilities of the value of the BI parameter \(\gamma\) and named this arbitrariness of the BI parameter \(\gamma\) as the BI parameter \(\gamma \) crisis [27 - 28]. 

Since, Barbero introduced this free parameter and Immirzi used it to explain canonical quantization method along with Regge calculus; the \(\gamma\) is known as the Barbero - Immirzi parameter. In short,  Barbero used one-parameter scale transformation to generalize the Ashtekar canonical transformation to a 
\(U(\gamma)\); While, Immirzi observed that such a transformation modifies the spectra of geometrical quantities of LQG [1 -5, 27-28].

\subsection{The Holst action and the BI parameter}

At present, there are two sort of version for the connection variables: \(SL(2,\mathbb{C})\) with a self duality of Yang-Mills sort of connection (i.e., Ashtekar connection) and the other is the connection with a real \(SU(2)\) (i.e., Barbero connection). In the second type of connection, the issue of the reality condition is not present. With the aid of the Holst action both of the connection can be obtained. The BI parameter \(\gamma\) is introduced in the Holst action as a multiplicative constant that governs the strength of the dual curvature correction. 

The Holst action generalizes the Hilbert-Palatini action using the BI parameter.  The Holst action can be derived in the following way using the Einstein-Hilbert action. In general relativity, the Einstein - Hilbert action is written as [1-5,15,29]
 
\begin{eqnarray}
S_{EH}(g_{\mu\nu}) = \frac{1}{16 \pi G} \int d^{4} x \sqrt{-g} g^{\mu\nu}R_{\mu\nu}
\end{eqnarray}

If \(e = \sqrt{-g}\) and \(8 \pi G = 1\) are taken; then [1-5,15,29], 
 
\begin{center}
\(S_{EH}(g_{\mu\nu}(e))= \int d^{4} x e e_{J}^{\mu} e^{\nu J} R_{\mu\nu\eta\tau} e_{K}^{\eta}e^{\tau K}\)
\end{center}

\begin{center}
\(= \int d^{4} x e e_{J}^{\mu} e_{K}^{\eta} F_{\mu\eta}^{JK} (\omega(e)) \)
\end{center}

Where, \(F_{\mu\eta}^{JK}(\omega(e)) = e^{J\mu} e^{\eta K} R_{\mu\nu\eta\tau} (e)\)

\begin{center}
\(S_{EH}(g_{\mu\nu}(e))= \int d^{4} x \frac{1}{4} \epsilon_{JKLM}\epsilon^{\mu\eta\alpha\beta} e_{\alpha}^{L}e_{\beta}^{M} F_{\mu\eta}^{JK}(\omega(e))\)
\end{center}

Hence, As a functional of densitized triad, the Einstein - Hilbert action takes the form [1-5,15,29], 

\begin{eqnarray}
\therefore S \left(e_{\mu}^{J}, \omega_{\mu}^{JK}\right) = \frac{1}{2}\epsilon_{JKLM} \int e^{J} \wedge e^{K} \wedge F^{LM}(\omega)
\end{eqnarray}

By considering the Palatini identity, i.e., \(\delta_{\omega} = F^{LM} (\omega) = d_{\omega} \delta_{\omega}^{LM}\) and taking the variation of equation (14); one gets [1-5,15,29]

\begin{center}
\(\delta_{\omega} S \left(e_{\mu}^{J}, \omega_{\mu}^{JK}\right) = \frac{1}{2}\epsilon_{JKLM} \int e^{J} \wedge e^{K} d_{\omega} \delta_{\omega}^{LM}\)
\end{center} 

\begin{eqnarray}
\therefore \delta_{\omega} S \left(e_{\mu}^{J}, \omega_{\mu}^{JK}\right) = - \frac{1}{2}\epsilon_{JKLM} \int d_{\omega} \left(e^{J} \wedge e^{K}\right) \wedge \delta_{\omega}^{LM}
\end{eqnarray}

If the coupling constant \(\frac{1}{\gamma}\) is added in the equation (14); then, one gets the Holst action [1-5,15,29]. i.e., 

\begin{eqnarray}
S(e, \omega) = \left(\frac{1}{2}\epsilon_{JKLM} + \frac{1}{\gamma}\delta_{JKLM}\right) \int e^{J} \wedge e^{K} \wedge F^{LM}(\omega)
\end{eqnarray}

Where \(\gamma\) is the BI parameter. The Holst action is equivalent to the Ashtekar Hamiltonian; if, \(\gamma = i\) is taken in the equation (16) [1-5,15,29].

In general, the Holst action is written as [1-5,15,29],

\begin{eqnarray}
S[e,A] = \frac{1}{8 \pi G} \left(\int d^4 x e e_{J}^{\mu}e_{K}^{\nu}F_{\mu\nu}^{JK} - \frac{1}{\gamma} \int d^4 x e e_{J}^{\mu}e_{K}^{\nu} *F_{\mu\nu}^{JK}\right)
\end{eqnarray}

Where, \(J, K...=0,1,2,3\) are internal indices, \(\mu, \nu... =0,1,2,3\) or \(c,d...=0,1,2,3\) (in some literature) are spatial indices, and * show self-duality within the equation (17) in the presence of the BI parameter \(\gamma\).  

\section{Various proposals on the physical significance of the BI parameter}

In this section, the historical time line of the research in the BI parameter and various proposals on the physical significance of the BI parameter are briefly given. After introducing each proposal, pros and cons of that proposal are given; in which the role of the BI parameter are explained.  Here, all proposals are explained only in context with the BI parameter \(\gamma\) up to relevant extent with the necessary mathematical treatment. 

\subsection{Historical timeline}

In the tabular form, the historical time line of research on the BI parameter in LQG is given in the chronological order. This table lists the enriched literature of the BI parameter and suggests why the study of the BI parameter in LQG is so important. This tabular listing of literature for the BI parameter implies that the BI parameter is itself a crucial research area in LQG. 

\begin{center}
\begin{stabular}{|p{0.05\textwidth} | p{0.07\textwidth} | p{0.73\textwidth}|}
\hline
\textbf{Sr. No.} & \textbf{Year} & \textbf{Research on  the BI parameter and its significance} \\
\hline
1 & 1986 & Discovery of Ashtekar variables \\
\hline
2 & 1995 & Real Ashtekar variables for Lorentzian signature space-times \\
\hline
3 & 1996 & Barbero's Hamiltonian derived from a generalized Hilbert-Palatini action \\
\hline
4 & 1996 & 	Black hole entropy from loop quantum gravity  \\
\hline
5 & 1996 & From Euclidean to Lorentzian General Relativity: The Real Way \\
\hline
6 & 1996 & Real and complex connections for canonical gravity \\
\hline
7 & 1997 & Quantum gravity and Regge calculus \\
\hline
8 & 1997 & 	Counting surface states in LQG \\
\hline
9 & 1997 & Immirzi parameter in quantum general relativity \\ 
\hline
10 & 1997 & On the constant that fixes the area spectrum in canonical quantum gravity \\
\hline
11 & 1998 & Quantum Geometry and Black Hole Entropy \\
\hline
12 & 2000 & Is Barbero’s Hamiltonian formulation a Gauge Theory of
Lorentzian Gravity? \\
\hline
13 & 2001 & Comment on ‘‘Immirzi parameter in quantum general relativity’’ \\
\hline
14 & 2003 & Quasinormal Modes, the Area Spectrum, and Black Hole Entropy \\
\hline
15 & 2004 & Black-hole entropy in loop quantum gravity \\
\hline
16 & 2004 & Black-hole entropy from quantum geometry \\
\hline
17 & 2005 & Origin of the Immirzi parameter \\
 \hline
18 & 2005 & Physical effects of the Immirzi parameter \\
\hline
19 & 2005 & On choice of connection in loop quantum gravity \\
\hline
20 & 2007 & On a covariant formulation of the Barbero–Immirzi connection \\
\hline
21 & 2007 & Renormalization and black hole entropy in Loop Quantum Gravity \\
\hline
22 & 2008 & From the Einstein-Cartan to the Ashtekar-Barbero canonical constraints, passing through the Nieh-Yan functional \\
\hline
23 & 2008 & The Barbero-Immirzi Parameter as a Scalar Field: K-Inflation from Loop Quantum Gravity? \\
\hline 
24 & 2008 & Topological Interpretation of Barbero-Immirzi Parameter \\
\hline
25 & 2009 & Peccei-Quinn Mechanism in Gravity and the Nature of the Barbero-Immirzi Parameter \\
 
26 & 2010 & A relation between the Barbero–Immirzi parameter and the standard model \\
\hline
27 & 2011 & Complex Ashtekar variables, the Kodama state and spinfoam gravity \\
\hline
28 & 2012 & The Quantum Gravity Immirzi Parameter—A General Physical
and Topological Interpretation \\
\hline
29 & 2012 & Complex Ashtekar Variables and Reality Conditions for Holst’s Action \\
\hline
30 & 2013 & Black Hole Entropy from complex Ashtekar variables \\
\hline
31 & 2014 & Geometric temperature and entropy of quantum isolated horizons \\
\hline
32 & 2014 & A Correction to the Immirizi Parameter of SU(2) Spin Networks \\
\hline
33 & 2014 & The Microcanonical Entropy of Quantum Isolated Horizon, `quantum hair' N and the Barbero-Immirzi parameter fixation \\
\hline
34 & 2015 & The holographic principle and the Immirzi parameter of loop
quantum gravity \\
\hline
35 & 2017 & Immirzi parameter without Immirzi ambiguity: Conformal loop
quantization of scalar-tensor gravity \\
\hline
36 & 2018 & Horizon entropy with loop quantum gravity methods \\
\hline
37 & 2018 & Generalizing the Kodama State I: Construction \\
\hline
38 & 2018 & Generalizing the Kodama State II: Properties and Physical Interpretation \\
\hline
39 & 2018 & Chiral vacuum fluctuations in quantum gravity \\
\hline
40 & 2018 & Black hole entropy from the SU(2)-invariant formulation of Type I isolated horizons \\
\hline
41 & 2018 & Black hole entropy and SU(2) Chern-Simons theory \\
\hline
42 & 2020 & On the value of the Immirzi parameter and the horizon entropy \\
\hline
\end{stabular}
\end{center}
\begin{center}
Table 1.1: Time line of research on the Barbero Immirzi parameter.
\end{center}

\subsection{The area operator and the BI parameter}

In LQG, the loop state as a graph or network \(\Theta\) with edges \(e_{i}\) denoted by elements of some gauge group. In general, this gauge group can be \(SU(2)\) or \(SL(2,\mathbb{C})\) [1-5, 12, 15, 17]. 

\begin{eqnarray}
\psi_{\Theta} = \psi (g_{1}, g_{2},...,g_{k})
\end{eqnarray}

Where, \(k = 0, 1, ,2.., n\) and \(g_{k}\) is the holonomy (group element) of connection \(A\) on the \(kth\) edge. In LQG, the spin network is used describe these loop states. Penrose [30-31] gave the notion of the spin network. In the spin network, the combinatorial principle of angular momentum is used and it defines the space-time in discrete way.  In LQG, the spin network is essential to represent the loop state [1-5, 12, 15, 17].  

The area of a 2D surface \(S\) that is embedded in any manifold \(\Sigma\) is defined as [1-5, 12, 15, 17]
 
\begin{eqnarray}
A_{S} = \int d^{2} x \sqrt{^{(2)}m}
\end{eqnarray}

Where, \(^{(2)}{m}\) is the determinant the metric \(^{(2)}m_{EF}\). The area is 2D; hence, the components of the 2D metric \(^{2)}m_{EF}\) can be denoted as the dyad basis \(e_{E}^{J}\), \(J,K\) internal indices  and \(E, F \in \lbrace x, y \rbrace\) are spatial indices [1-5, 12, 15, 17]. Thus,

\begin{eqnarray}
^{2}m_{EF} = e_{E}^{J}e_{F}^{K}\delta_{JK}
\end{eqnarray}

The determinant of \(^{2}m_{EF}\) can be written as [1-5, 12, 15, 17] 

\begin{eqnarray}
det \left(^{2}m_{EF}\right) = m_{11}m_{22} - m_{12}m_{21} = \vec{e}_{z}\cdot\vec{e}_{z}
\end{eqnarray}

Hence, the equation (19) becomes [1-5, 12, 15, 17], 

\begin{eqnarray}
A_{S} = \int d^{2} x \sqrt{\vec{e}_{z}\cdot\vec{e}_{z}}
\end{eqnarray}

In LQG, the frame field \(e_{E}^{k}\) and the connection \(A_{k}^{E}\) are conjugates.    For instance, \(e_{E}^{k} \rightarrow - i \hbar \frac{\delta}{\delta A_{k}^{E}}\) [1-5, 12, 15, 17]. By putting this value into the equation (22),

\begin{figure}
\begin{center}
\includegraphics[scale=0.5]{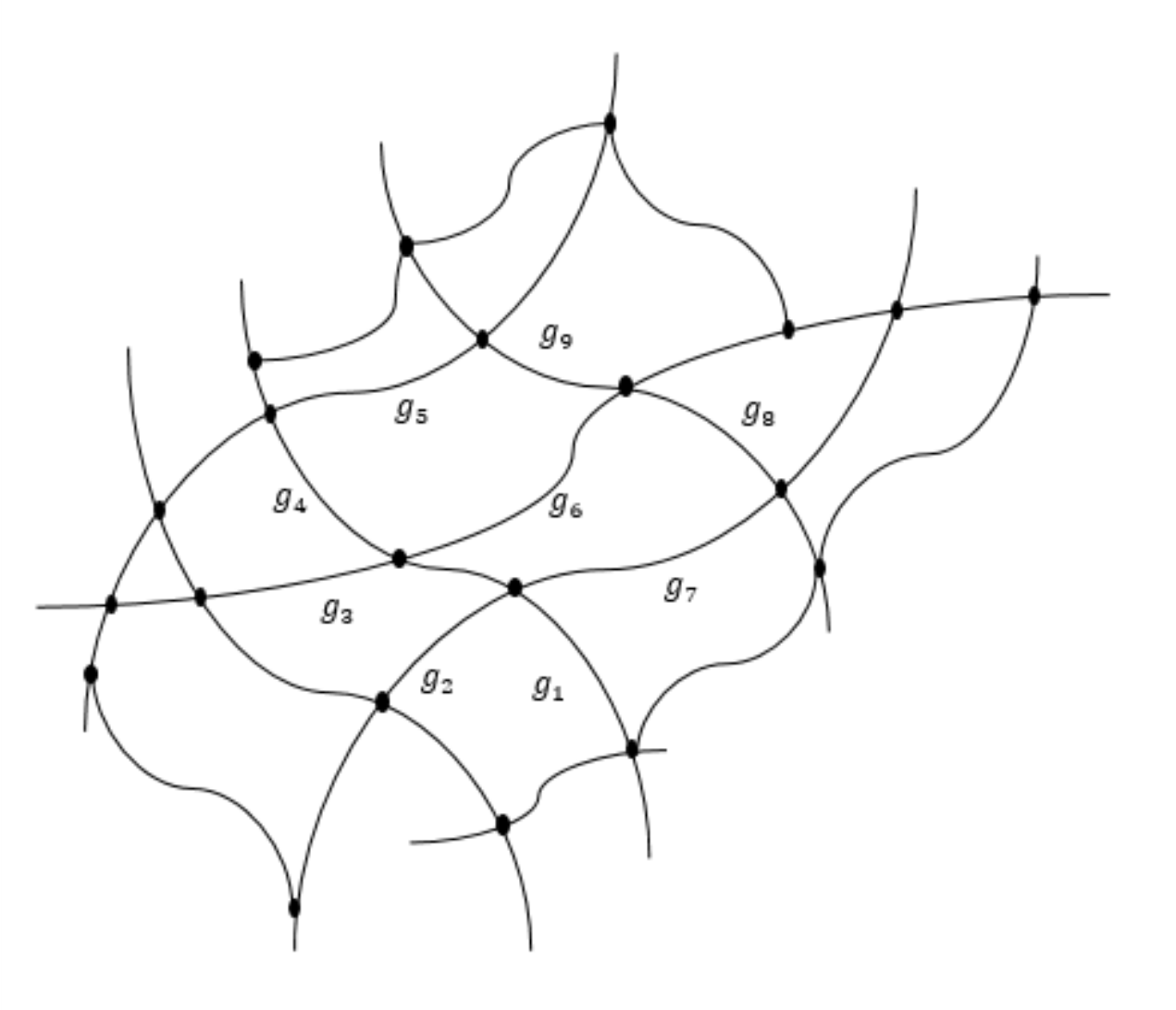}
\end{center}
\begin{center}
Fig. 1: A diagram of spin network.
\end{center}
\end{figure}

\begin{eqnarray}
\hat{A}_{S} = \int d^{2} x \sqrt{\delta_{JK} \frac{\delta }{\delta A_{J}^{z}}\frac{\delta }{\delta A_{K}^{z}}}
\end{eqnarray}

In LQG, \(e_{E}^{J}e_{F}^{K} = \frac{\delta}{\delta A_{J}^{E}}\frac{\delta}{\delta A_{K}^{F}} = n_{E}n_{F} J^{J}J^{K}\). For \(SO(3)\) group, the generator is the angular momentum operator \(J^{J}\). Here \(n_{E}\) and \(n_{F}\) are unit tangent vector. therefore, the equation of the area operator becomes [1-5, 12, 15, 17]

\begin{eqnarray}
\hat{A}_{S} = \Sigma_{p} \sqrt{\delta^{JK} n_{E}n_{F} J^{J}J^{K}} 
\end{eqnarray}

\begin{center}
\( = \Sigma_{p} \sqrt{\delta^{JK} \hat{J}_{J}\hat{J}_{K}}\Psi_{\Theta} (\because n^{c}n_{c} = 1)\)
\end{center}

\begin{eqnarray}
\therefore \hat{A}_{S}  = \Sigma_{p} \sqrt{\mathbf{J^{2}} } 
\end{eqnarray}

But, the quantum state of \(J\) is \(\mathbf{J^{2}} \vert j\rangle = \hbar^{2} j(j+1)\vert j\rangle\); hence, the equation is [1-5, 12, 15, 17]

\begin{eqnarray}
\hat{A}_{S} \Psi_{\Theta} = l_{P}^{2} \Sigma_{p} \sqrt{j_{p}(j_{p} + 1)} \Psi_{\Theta}
\end{eqnarray}

Where, \(l_{P}^{2} = \frac{G \hbar}{c^{3}}\) is the Planck area.

In LQG, lines of spin network are allowed to intersect. Any surface \(\Sigma\) acquires area through the puncture of these lines [1-5, 12, 15, 17].

 \begin{figure}
\begin{center}
\includegraphics[scale=0.5]{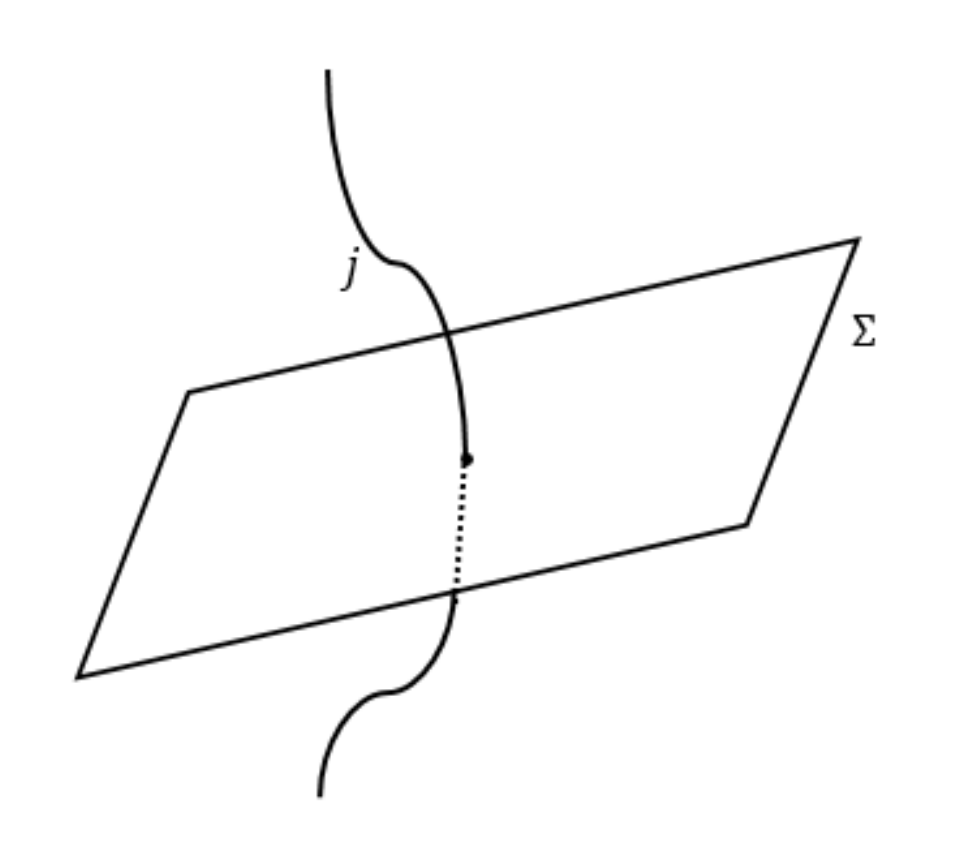}
\end{center}
\begin{center}
Fig. 2: A diagram of surface puncture.
\end{center}
\end{figure}

In standard form, the area operator with the \(\gamma\) can also be written as [1-5, 12, 15, 17]

\begin{eqnarray}
\therefore \hat{A}_{S}  = \gamma l_{P}^{2} \Sigma_{p} \sqrt{j_{p}(j_{p} + 1)} 
\end{eqnarray}

The proportionality coefficient in the formula of area operator in LQG includes the BI parameter \(\gamma\) [1-5, 23]. In 1998, Krasnov [32] found that the multiplicative factor of the area operator is \(8\pi\gamma\). Hence, the equation is [1-5, 12, 15, 17]

\begin{eqnarray}
A = 8 \pi \gamma l_{P}^{2} \sqrt{j(j+1)}
\end{eqnarray}

Similar to the area operator, the BI parameter also appears in the volume operator. The spectrum of volume operator can only be understood; if, the BI  parameter \(\gamma\) is real valued [1-5]. 

\subsubsection{Pros and cons of the area operator and the BI parameter}

The spectrum of the area and the volume operator and its eigenvalue can only be understood with the real valued \(\gamma\). As mentioned, the complex valued \(\gamma\) makes these operators complex valued and the significance of these complex valued operators are ambiguous. Is there any valid significance of the area operator and the volume operator with the complex valued \(\gamma\)? This question is still open.

\subsection{The BI parameter and black hole entropy calculation in LQG}

The expression of entropy of black hole in Planck units calculated semi classically by Hawking is written as [1 - 5, 23], 

\begin{eqnarray}
S = \frac{A}{4}
\end{eqnarray}

In 1996, Rovelli [33] calculated black hole entropy within LQG using the statistical framework. 

In LQG, any surface obtains area when link of spin network punctures that surface. One can allots micro states to each puncturing to surface. Thus, the micro states are associated to the discrete pieces of the surface that provides the value of area spectrum by puncturing. So, the entropy \(S\) is proportional to the log of the number of ways in which the sphere can be punctured that provides an area within each macroscopic interval [33]

If the eigenvalue value of the area operator with the  \(j_{p}\) of the form \(\frac{m_{p}}{2}\), (where, \(m_{p} \in Z\)) is written as [33]

\begin{eqnarray}
A_{p} = 4 \pi \gamma l_{P}^{2} \sqrt{m_{p}(m_{p} + 2)}
\end{eqnarray}

For an interval \([A + \delta A, A - \delta A]\), where, \(\delta A\) is some small interval \((\frac{\delta A}{A} \ll 1)\) \(A\) is a macroscopic value of area and  the allowed  number \(N(M)\) of sequences of integers \( \lbrace m_{p},...,m_{N}\rbrace\)  \(p=1,2,3,…N\) is determined by the number \(N\) of edges that puncture the surface , so that the determined value for the total area lies within the given interval \(M = \frac{A}{4 \pi \gamma l_{P}^{2}}\) [33]

The number of sequences \(N(M)\), in which each sequence is \(\lbrace m_{p}\rbrace\), can be given as [33]

\begin{eqnarray}
M = \frac{A}{4 \pi \gamma l_{P}^{2}} = \Sigma_{p} \sqrt{m_{p} (m_{p} + 2)}
\end{eqnarray}

The number of sequences are indicated by \(N_{+}(M)\) such that \(\Sigma_{p} m_{p} = M\) and the number of sequences are indicated by \(N_{-}(M)\) such that \(\Sigma_{p}(m_{p}+1) = M\). Hence, the given set of inequalities implies that [33]

\begin{eqnarray}
N_{-}(M) < N(M) < N_{+}(M)
\end{eqnarray}

From calculation, \(ln N_{+}(M) = ln 2\), \(ln N_{-}(M) = ln \frac{1 + \sqrt{5}}{2}\) and \(ln N(M) = d M\). From equation (32), The inequalities is now given as [33],  

\begin{center}
\(ln \frac{1 + \sqrt{5}}{2} < d < ln 2\)
\end{center}

\begin{eqnarray}
0.48 < d < 0.69
\end{eqnarray}

By taking, \(M = \frac{A}{8 \pi \hbar G}\); one gets

\begin{eqnarray}
ln N(A) = d \frac{A}{8 \pi \hbar G}
\end{eqnarray}

\begin{center}
\(S (A) = k ln N(A)\)
\end{center}

\begin{eqnarray}
\therefore S(A) = c \frac{k}{\hbar G} A
\end{eqnarray}

Where, \(c = \frac{d}{8 \pi} = \frac{1}{4}\) is constant 

In 1997, Ashtekar et al [34] showed that spin networks explains spacetime geometry outside a black hole. Some edges of this spin network puncture the event horizon, provides the value of area through this contribution. The \(U(1)\) Chern - Simons theory explains the quantum geometry of the horizon.  In this formalism, the rotation of \(SO(2)\) describes \(2D\) geometry, that is isomorphic to \(U(1)\). The entropy of black hole is calculated through counting of spin network states relevant to an event horizon. Thus, the expression of black hole entropy in LQG is [1 - 5, 34],

\begin{eqnarray}
S = \frac{\gamma_{0}A}{4\gamma} 
\end{eqnarray}

There are two possibilities for the value of \(\gamma_{0}\)[1-5, 34]; i.e.,

\begin{eqnarray}
\gamma_{0} = \frac{ln (2)}{\sqrt{3} \pi} 
\end{eqnarray}

or

\begin{eqnarray}
\gamma_{0} = \frac{ln (3)}{\sqrt{8} \pi}
\end{eqnarray}

 The value of \(\gamma_{0}\) relies on the choice of the gauge group. By taking \(\gamma_{0} = \gamma\), one gets actual black hole entropy formula calculated by Hawking [1 - 5, 34] i.e. 

\begin{eqnarray}
S =\frac{\gamma_{0} A}{4\gamma_{0}} = \frac{A}{4}
\end{eqnarray}

 \begin{figure}
\begin{center}
\includegraphics[scale=0.5]{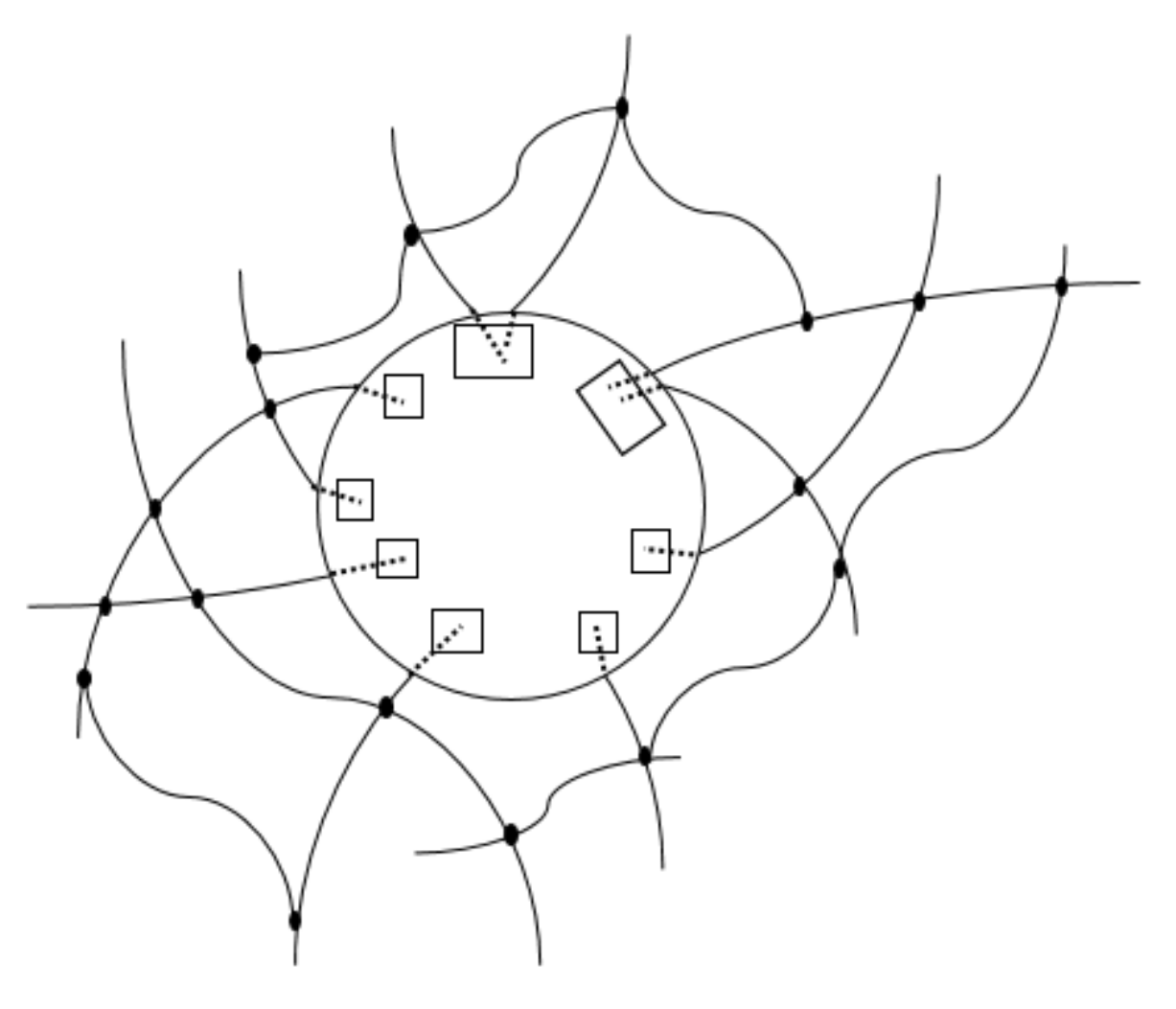}
\end{center}
\begin{center}
Fig. 3: The black hole entropy through puncture in the surface of the event horizon.
\end{center}
\end{figure}

This calculation is true for each sort of black hole. The black hole entropy calculation in LQG is quite enriched research area. The value of the BI parameter \(\gamma\) and black hole entropy formula in LQG is a topic whose implications are far - reaching  [1 - 5, 34].    

In 2002, Dreyer [35] Fixed the value of BI parameter \(\gamma\) using classical quasinormal mode spectrum of a black hole and gave black hole entropy formula in LQG with \(SO(3)\) group instead of \(SU(2)\). Instead of \(j_{min}=\frac{1}{2}\); the value of \(j_{min} = 1\) is taken in the expression of area operator. i.e.,  

\begin{eqnarray}
\Delta A = A(j_{min}) = 8 \pi \gamma l_{P}^{2} \sqrt{j_{min}(j_{min}+1)}
\end{eqnarray} 

In this case, the change in the mass due to the frequency of the quasi normal modes is [35], 

\begin{eqnarray}
\Delta M = \hbar \omega_{QNM} = \frac{\hbar ln 3}{8 \pi M}
\end{eqnarray}

For a Schwarzschild black hole, the area and the mass are related to each other by the relation \(A = 16 \pi M^{2}\). Change in area corresponding to the mass change is given as [35] 

\begin{eqnarray}
\Delta A = 4 ln 3l_{P}^{2}
\end{eqnarray}  

The expression for the BI parameter \(\gamma\) is obtained through comparison between the equation (40) and (42) [35],

\begin{eqnarray}
\gamma = \frac{ln 3}{2 \pi \sqrt{j_{min}(j_{min}+1)}}
\end{eqnarray} 

If \(j_{min} =1\) is taken in the equation (43); then,  the fixed value of the BI parameter is [35],

\begin{eqnarray}
\gamma = \frac{ln 3}{2 \pi \sqrt{2}}
\end{eqnarray}

Thereafter, Meissner [36] fixed the value of area in LQG by fixing \(\gamma_{M}\) and \(\gamma\) by comparing the Bekenstein-Hawking entropy formula with the derived formula. The equation of Bekenstein - Hawking entropy formula is 

\begin{eqnarray}
S = \frac{1}{4}\frac{A}{l_{P}^{2}}
\end{eqnarray}

Here, the derived expression of the black hole entropy formula is [36] 

\begin{eqnarray}
S = ln N(a) = \frac{\gamma_{M}}{4\gamma}\frac{A}{l_{P}^{2}} + O ln(A)
\end{eqnarray}

Here, by comparing the derived black hole entropy formula with the Bekenstein Hawking formula; one gets [1 - 5, 36].

\begin{eqnarray}
\gamma = \gamma_{M}
\end{eqnarray} 

The calculated value of \(\gamma_{M}\) is, 

\begin{eqnarray}
\gamma_{M} = 0.2375...
\end{eqnarray}

or

\begin{eqnarray}
\gamma_{M} = 0.2739...
\end{eqnarray}

In 2004, Domagala and Lewandowski [37] gave microscopic degrees of freedom to count the black-hole entropy. On the basis of a ratio i.e., \(\frac{ln N(a)}{a}\), for large \(a\), the value of entropy is 

\begin{eqnarray}
\frac{ln 2}{4\pi \gamma l_{P}^{2}} a \leq ln N(a) \leq \frac{ln 3}{4\pi \gamma l_{P}^{2}} a
\end{eqnarray}

Hence, the upper bound and lower bound value for \(\gamma\) is [37],

\begin{eqnarray}
\frac{ln 2}{\pi} \leq \gamma \leq \frac{ln 3 }{\pi}
\end{eqnarray} 

Since, the spin greater then \(\frac{1}{2}\)  also contributes to the entropy; this contribution are also considered here [37]. 

In 2007, Jacobson [38] studied renormalization and black hole entropy in LQG. For black holes, he found that the microscopic state counting are related to the Newton's universal constant \(G\).  The equation of this study is given as

\begin{eqnarray}
S_{LQG} = \frac{b}{\gamma} \frac{A}{\hbar G}
\end{eqnarray}

Where, \(b\) is a numerical constant. In LQG, from the equation (52), one can see that the entropy is related to the area of the horizon \(A\) \((S_{LQG} \propto A)\) and the gravitational constant \(G\) \((S_{LQG} \propto \frac{1}{G})\) [38].

He found that one should compare this formula with the actual Bekenstein - Hawking entropy formula after accounting the scale dependence of Newton's constant and area. For any value of the BI parameter \(\gamma\), if some property of renormalization is followed then one can compare both entropy formulas. The BI parameter should be of the order of \(\gamma = 4b\) to match the black hole entropy in LQG with the Bekenstein-Hawking formula [38].

In 2013, Frodden et al [39] found that by taking complex valued Ashtekar variables, black hole entropy   formula in certain condition is achieved. In this case the BI parameter can be complex valued \((\gamma = \pm i)\).  This paper shows that the number of micro states \(N_{\Gamma}\left(A, \gamma \rightarrow \pm i \right)\) acts as \(exp\left(\frac{A}{(4l_{P}^{2})}\right)\) for certain case i.e. for large area \(A\) in the large spin semi classical limit. With  respect to the complex self-dual Ashtekar
connection, \(N_{\Gamma}\left(A, \pm i \right)\)  is the number of states for a theory to be defined. 

The \(SU(2)\) Chern-Simons theory is related to the study of the black hole entropy in LQG. From \(SU(2)\) to  \(SL(2, \mathbb{C})\) representation, the formula for the Hilbert space of \(SU(2)\) Chern-Simons theory follows a specific analytic continuation with constraints of self-duality. The complex formulation ( with the Ashtekar variables) within this proposal , gives the derivation of the black hole entropy within LQG formalism for large spin asymptotic domain which is semi classical in nature [39]. Hence, 

\begin{eqnarray}
log \left(N_{\Gamma}(A, \pm i)\right) \sim \frac{A}{4 l_{P}^{2}}
\end{eqnarray}
 
There are many other papers  available on black hole entropy in LQG [40-48].

\subsubsection{Pros and cons of the BI parameter and black hole entropy calculation in LQG}

On the basis of the Black hole entropy calculation in LQG, one gets many expressions for the real valued BI parameter \(\gamma\) such as \(\gamma_{0} = \frac{ln (2)}{\sqrt{3} \pi}\), \(\gamma_{0} = \frac{ln (3)}{\sqrt{8} \pi}\) and \(\gamma = \frac{ln 3}{2 \pi \sqrt{2}}\). The numerical value for the BI parameter \(\gamma\) is either \(0.2375...\) or \(0.2739...\) based on the calculation. The BI parameter is also  expressed in terms of a numerical constant \(b\) i.e., \(\gamma = 4b\). With the complex valued BI parameter \(\gamma\), the black hole entropy can also be calculated using \(SU(2)\) Chern-Simons theory. Whether it is the real valued or the complex value BI parameter \(\gamma\); the black hole entropy can be calculated in LQG. However, the interpretation of the complex valued BI parameter within the black hole entropy formula in LQG is difficult to comprehend.  

\subsection{The BI parameter as Immirzi ambiguity}

In LQG, geometrical obeservable such as area and volume are quantized and they exhibits discrete spectrum.  In 1996, Immirzi noticed that LQG does not determine the complete scale of such spectra. Immrizi also observed that one can have different spectra for the same geometrical quantities if he begins with scaled elementary variables. The algebra of holonomy relies on a free parameter that gives family of one parameter of quantum theories with inequivalance. The BI parameter \(\gamma\) is this family of one parameter [49].

There is a certain symmetry under study according to which classical theory is identified as a canonical transformation; but, one can not identify it as a unitary transformation of quantum theory. Since, holonomy is  operator of LQG and because of weird sort of representation of LQG one has to consider the BI parameter \(\gamma\) as ambiguity [49].

In LQG, there are two connection  i.e., \(A\) and \(\Gamma\). Therefore, one has to create the \(\gamma\) scaled connection namely, \(A_{\gamma}\) by interpolation between different connections. Thus the elementary excitation of LQG namely Wilson loop of \(A_{\gamma}\) results into different for various value of the \(\gamma\). Therefore, some physical spectrum of quantity of LQG relies on the \(\gamma\) [49].

Another thing is metric information resides in the \(E\)(conjugate variable). Since, \(E\) is conjugate to connection; in quantum formalism, it is written as derivative operators that acts on functions over the group. Over the group manifold, any geometrical quantity that is function of \(E\) behaves as elliptic operator that results into discrete spectrum. Such elliptic operators possess non-vanishing scalar dimension relative to the affine scaling of the connection. Hence, in the elliptic geometric operators spectrum, ambiguity is introduced i.e., the Immirzi ambiguity. This ambiguity influences discreteness of space in LQG. The authors also described that various interpretation regarding the BI parameter \(\gamma\) are incorrect in context with the Immirzi ambiguity. The authors also gave various models such as 'harmonic oscillator with no Immirzi ambiguity', 'particle on a circle with no Immirzi ambiguity' and 'a simple model with \(\gamma\) as a free parameter'. Due to lake of space, here only cause of the Immirzi ambiguity and its effect in LQG are given   [49]. 

In year 2001, Samual [50] commented that Interpretations as  the Immirzi ambiguity is not clear and does not give any agreement on its origin and significance. All interpretations of the BI parameter as Immirzi ambiguity, seems unclear and does not give any satisfactory explanation about origin and significance of it. Moreover, the examples of the Immirzi ambiguity are not  not real but artificial generated through the compactification of the configuration space. 

In 2017, Veraguth and Wang [51] gave a proposal in which they explained LQG without Immirzi ambiguity using conformal LQG. The conformal LQG gives a way to loop quantization through conformally equivalent class of metrics. The conformal geometry gives an extended symmetry to permit a reformulated BI parameter. In scalar-tensor gravity, this can be done by conformal frame transformations. In this proposal, the authors showed that the LQG along with a conformally transformed Einstein metric having unlike values of the relevant BI parameter are connected by a conformal frame with global change. The conformal LQG is free from the Immirzi ambiguity. they defined the Ashtekar variables in the following way,

\begin{eqnarray}
\left(A_{c}^{'j} = \Gamma_{c}^{j} + \gamma \kappa K_{c}^{j}, \frac{1}{\gamma \kappa} E_{c}^{j}\right)
\end{eqnarray}

Here, \(\gamma\) is the conformal BI parameter.

\subsubsection{Pros and cons of the BI parameter as Immirzi ambiguity}

Research on the Immirzi ambiguity is still incomplete. As mentioned, there is a speific symmetry under study according to which classical theory is identified as a canonical transformation; but, one can not identify it as a unitary transformation of quantum theory. This is the reason behind the Immirzi ambiguity. One parameter family of the BI parameter is the another reason. The \(\gamma\) scaled connection, namely, \(A_{\gamma}\) for various values of the BI parameter is also different. The elliptic nature of the geometric operator due to the frame field \(E\) that is conjugate to the connection \(A\) also responsible for the Immirzi ambiguity. However, the conformal formalism of LQG may remove the Immirzi ambiguity.
 
\subsection{Origin of the BI parameter}

In 2005, Chou et al. [52] found a technique through which a ratio is obtained which equals the BI parameter \(\gamma\). They used quadratic spinor techniques in which the physical significance and effect of the BI parameter \(\gamma\) becomes obvious in general relativity. The authors also inferred that without other matter field in GR; the BI parameter \(\gamma\) as a observable is a physical property of sector of gravity. 

Firstly, they defined the Holst action in novel way [52] i.e.,

\begin{eqnarray}
S[\mathbf{e}, \omega] = \alpha \int * \left(\mathbf{e}^{c} \wedge \mathbf{e}^{d} \right) \wedge R_{cd}(\omega) + \gamma \left(\mathbf{e}^{c} \wedge \mathbf{e}^{d} \right) \wedge R_{cd}(\omega)
\end{eqnarray} 

Where, the BI parameter \(\gamma\) is a ratio i.e., \(\gamma = \frac{\alpha}{\beta}\), \(c, d... = 0,1,2,3\) and * shows duality.

Thereafter, the authors compared the equation (55) with the quadratic spinor Lagrangian [52] i.e., 

\begin{eqnarray}
\mathcal{L}_{\psi} = 2D \left( \bar{\psi} \mathbf{e}\right)\gamma_{5}D\left( \psi \mathbf{e}\right)
\end{eqnarray}

Where, \(\psi\) is auxiliary spinor field, \(\mathbf{e}\) is \(\mathbf{e} = \mathbf{e}^{c}\gamma_{c}\), \(D\) is covariant derivative and \(\gamma_{c}\) is the Dirac gamma matrices.

By defining the spinor curvature identity [52], i.e., 

\begin{eqnarray}
\begin{aligned}
2D \left( \bar{\psi} \mathbf{e}\right)\gamma_{5}D\left(\mathbf{e} \psi \right) = & \ \bar{\psi}\psi R_{cd} \wedge *\left(\mathbf{e}^{c} \wedge \mathbf{e}^{d}\right) \\ & + \bar{\psi}\gamma_{5}\psi R_{cd} \wedge \mathbf{e}^{c} \wedge \mathbf{e}^{d}  + d [D \left(\bar{\psi} \mathbf{e}\right) \gamma_{5} \mathbf{e} \psi + \bar{\psi}\mathbf{e}\gamma_{5} D \left(\mathbf{e} \psi\right) ]
\end{aligned}
\end{eqnarray}

For \(\omega [e]\), the equation of motion is [52], 

\begin{eqnarray}
D [\bar{\psi}\psi *\left(\mathbf{e}^{c} \wedge \mathbf{e}^{d}\right) + \bar{\psi}\gamma_{5}\psi \left(\mathbf{e}^{c} \wedge \mathbf{e}^{d}\right)] = 0
\end{eqnarray}

They found that the BI parameter \(\gamma\) can be written as a ratio of scalar and pseudoscalar contributions in the theory [52] i.e., 

\begin{eqnarray}
\gamma = \frac{\langle \bar{\psi}\psi\rangle}{\langle\bar{\psi}\gamma_{5}\psi\rangle}
\end{eqnarray}

If \(\bar{\psi}\psi = 1\) and  \(\bar{\psi}\gamma_{5}\psi = 0\); then, \(\gamma = \infty\). The BI parameter \(\gamma = i\) corresponds to Ashtekar formalism with self duality; while, \(\gamma = 1\) satisfies the action of the Hamiltonian given by Barbero. Therefore, the BI parameter \(\gamma\) implies that how Einstein gravity can be distinguished from the other gravitation theory with general covariance. In other words, this ratio can be seen as a measure of how gravity differs from covariant gravity. Such a techniques permits the renormalization scale \(\mu\) regarding the BI parameter \(\gamma\) via spinor's expectation value in quantization process \((\therefore \langle\bar{\psi}\psi\rangle_{\mu}, \langle\bar{\psi}\gamma_{5}\psi \rangle_{\mu})\). Here, \(\bar{\psi}\gamma_{5}\psi\) is not real function. To get \(\bar{\psi}\gamma_{5}\psi\) to be real, one has to use anti-commuting spinor to achieve real Ashtekar variables [52].

\subsubsection{Pros and cons of origin of the BI parameter}

This proposal gives the origin of the BI parameter \(\gamma\) using quadratic spinor techniques in which  a ratio of scalar and pseudoscalar contributions is defined as the BI parameter \(\gamma\). The anti-commuting spinor is necessary to get real valued \(\bar{\psi}\gamma_{5}\psi\) and the Ashtekar variables. 

\subsection{On a covariant formulation of the BI connection}

In 2007, Fatibene et al [53] gave a proposal on covariant formulation of the BI connection in which they defined a global covariant \(SU(2)\) - connection over whole spacetime that limits to general the Barbero - Immirzi connection on a given slice of space. The BI connection is a collective \(SU(2)\) gauge connection on a \(3D\) surface \(S\subset \mathcal{M}\) in \(4D\) spacetime \(\mathcal{M}\). On the basis of groups and spacetime involved in the theory,   the BI connection is global. But, the \(SU(2)\) principal bundles \( ^{+}\Sigma\) over one \(3D\) base \(S\) should be trivial. In this paper, the global aspects of the BI connection, the covariant formulation of the BI connection with its spacetime interpretation and the Lorentzian case are investigated. Here, the point of study is the BI connection. The BI parameter \(\gamma\) is less emphasized in this study.
Thus, this investigation is covered with necessary details.

\subsubsection{Pros and cons of a covariant formulation of the BI connection}

This proposal advocates usual interpretation of the BI parameter \(\gamma\) (real valued) on the basis of the black hole entropy calculation in LQG.

\subsection{The BI parameter as a scalar field}

In 2008, Taveras and Yunes [54] gave a proposal on the BI parameter \(\gamma\) as a scalar field. They studied LQG based generalization of general relativity in which they modified the Holst action.

The authors, scalarize the BI parameter \(\gamma\) in the Holst action. It means that the BI parameter \(\gamma\) is promoted as  a field under the integral of the dual curvature term. In this formalism, the BI parameter \(\gamma\) acts as a dynamical scalar field. This formalism gives a non-zero torsion tensor which modifies the field equations through quadratic first derivatives of the BI field. Such a modification is similar to general theory of relativity with non-trivial kinetic energy in the presence of a scalar field [54].  

Before promoting the BI parameter \(\gamma\), the author firstly modified the Holst action [54]. i.e.,

\begin{eqnarray}
S = \frac{1}{4\kappa} \int \epsilon_{JKLM} e^{J} \wedge e^{K} \wedge F^{LM} + \frac{1}{2\kappa} \int \bar{\gamma}e^{J} \wedge e^{K} \wedge F_{LM} + S_{mat}
\end{eqnarray} 

Where, \(\kappa = 8\pi G\), The coupling field \(\bar{\gamma}\) is \(\bar{\gamma} = \frac{1}{\gamma}\), \(\varepsilon_{JKLM}\) is Levi-Civita tensor and \(S_{mat}\) is the action for additional matter degrees of freedom. To introduce torsion and contorsion in the equation (60), one can simplifies the Holst action in the following way [54].

\begin{eqnarray}
S = \frac{1}{4\kappa} \int \epsilon_{JKLM} e^{J} \wedge e^{K} \wedge e^{Q} \wedge e^{R} \frac{1}{2} F_{QR}^{LM} + \frac{1}{2\kappa} \int \bar{\gamma}e^{J} \wedge e^{K} \wedge e^{L} \wedge e^{M} \frac{1}{2} F_{JKLM} + S_{mat}
\end{eqnarray}

\begin{center}

\( = \frac{1}{8\kappa} \int \epsilon_{JKLM} (-\tilde{\sigma}) \epsilon^{JKQR} F_{QR}^{LM} + \frac{1}{4\kappa} \int \bar{\gamma} (-\tilde{\sigma})  \epsilon^{JKLM}   F_{JKLM} + S_{mat}\)
\end{center}

By simplification, one gets [54]  

\begin{eqnarray}
\therefore S = \frac{1}{2\kappa} \int \tilde{\sigma} \left[\delta_{LM}^{[QR]} F_{QR}^{LM} - \frac{\bar{\gamma}}{2} \epsilon^{JKLM}F_{JKLM}\right] + S_{mat}
\end{eqnarray}

Where, \(\tilde{\sigma} = d^{4}x e = d^{4}x \sqrt{-g}\), and \(e^{J} \wedge e^{K} \wedge e^{L} \wedge e^{M} = - \tilde{\sigma} \epsilon^{JKLM} \).

In the simpler form, the modified form of the Holst action is [54],

\begin{eqnarray}
S = \frac{1}{2\kappa} \int d^{4} x e p_{LM}^{JK} e_{J}^{\mu} e_{K}^{\nu}F_{\mu\nu}^{LM}
\end{eqnarray}

where, \(p_{LM}^{JK} = \delta_{J}^{[L}\delta_{K}^{M]} - \frac{\bar{\gamma}}{2} \epsilon_{JK}^{LM}\). 

Thereafter, the authors gave field equations  with the modified Holst action and its solutions. They also gave effective action and inflation with the BI parameter \(\gamma\) as a dynamical scalar field [54].   

 In 2009,  Calcagni and  Mercuri [54] also promoted  the BI parameter \(\gamma\) as a field in the canonical formalism of pure gravity. In this paper, The authors investigated the parity properties of the field of the BI parameter \(\gamma\) by performing the decomposition of torsion into irreducible components. Under local Lorentz group, they suggested that the BI parameter \(\gamma\) is ought to be pseudoscalar to conserve the transformation properties of these components.
 
To understand the Riemann - Cartan space-time, one has to generalize the Holst formalism. This can be done by adding a torsion part in the Holst action. It gives net coupling with the BI parameter \(\gamma\) that gives rise to the Nieh - Yan density [54].

The field of BI parameter \(\gamma\) is a real canonical pseudoscalar field; for \(\gamma = \gamma(x)\) coupled with the Nieh - Yan invariant. The BI parameter \(\gamma\) is pseudoscalar in nature because of the axial component of torsion  which is proportional to the partial derivative of the BI parameter \(\gamma\). In the absence of matter, the authors studies the field of the BI parameter \(\gamma\) in the first-order Hamiltonian formalism. Here, the derivation of the action in Lagrangian formalism is avoided; since, the subject of study is the field of the BI parameter in Hamiltonian formalism (canonical).   The authors also compared the Holst case with the Nieh - Yan case. The total Hamiltonian in the form of the action in the Holst case is expressed as [54] 

\begin{eqnarray}
H_{D} = \int d^{3} x \left(\Lambda^{j}\mathcal{R}_{j} + N^{\beta}\mathcal{H}_{\beta} + N\mathcal{H}\right)
\end{eqnarray}

Where, \(\Lambda^{j}\), \(N^{\beta}\) and \(N\) are Lagrange undetermined multipliers. \(\mathcal{R}_{j}\), \({H}_{\beta}\) and \(\mathcal{H}\) are the rotation, super momentum, and super Hamiltonian constraints respectively.

The expression of all constraints are written as [54],

\begin{eqnarray}
\mathcal{R}_{j}  \equiv \epsilon_{jk}^{l} K_{\beta}^{k}E_{l}^{\beta} 
\end{eqnarray}

\begin{eqnarray}
\mathcal{H}_{\beta} \equiv E_{j}^{\eta} D_{[\beta}K_{\eta]}^{j} + \Pi \partial_{\beta}\gamma \approx 0
\end{eqnarray}

\begin{eqnarray}
\begin{aligned}
\mathcal{H} \equiv & \ - \frac{1}{2 e} E_{j}^{\beta}E_{k}^{\eta} \left(\epsilon_{l}^{jk}R_{\beta\eta}^{l}  + 2 K_{[\beta}^{j}K_{\eta]}^{k}\right) \\ & + \frac{1+\gamma^{2}}{3 e} \Pi^{2} - \frac{3}{4} \frac{e}{1+\gamma^{2}}\partial_{\beta} \gamma \partial^{\beta} \gamma \approx 0
\end{aligned}
\end{eqnarray}

For the Nieh - Yan case, the super Hamiltonian is denoted as [54] \(\mathcal{H}\). i.e., 

\begin{eqnarray}
\begin{aligned}
\mathcal{H} \equiv & \ - \frac{1}{2 e} E_{j}^{\beta}E_{k}^{\eta} \left(\epsilon_{l}^{jk}R_{\beta\eta}^{l}  + 2 K_{[\beta}^{j}K_{\eta]}^{k}\right) \\ & + \frac{1}{3 e} \Pi^{2} - \frac{3}{4} e \partial_{\beta} \gamma \partial^{\beta} \gamma \approx 0
\end{aligned}
\end{eqnarray}

Here, in the canonical formalism, the factor \(\left(1+\gamma^{2}\right)\) is disappeared in the contribution of the pseudo scalar field for the Nieh - Yan term. The Nieh - Yan term exhibits a shift symmetry i.e., \(\gamma \rightarrow \gamma +\gamma_{0}\) [54].

\subsubsection{Pros and cons of the BI parameter as a scalar field} 

Taveras and Yunes defined the BI parameter \(\gamma\) as a dynamical scale field in the Holst action with non zero torsion tensor; while, Calcagni and  Mercuri defined the BI parameter \(\gamma\) as a field in the canonical formalism. These proposals provide new significance of the BI parameter; however, the BI parameter \(\gamma\) is sometimes complex - valued is not still clear. 
 
\subsection{Topological interpretation of the BI parameter}

In 2008, Date et al[55] gave a proposal on topological interpretation of the BI parameter \(\gamma\).

In terms of the Holst formalism, the Hilbert - Palatini Lagrangian as the Lagrangian density can be written as [55], 

\begin{eqnarray}
\mathcal{L} = \frac{1}{2} e \Sigma_{JK}^{\mu\nu} R_{\mu\nu}^{JK}(\omega) + \frac{\gamma}{2} e  \Sigma_{JK}^{\mu\nu} \tilde{R}_{\mu\nu}^{JK}(\omega) 
\end{eqnarray}

Where, \(\Sigma_{JK}^{\mu\nu} := \frac{1}{2} \left(e_{J}^{\mu}e_{K}^{\nu} - e_{K}^{\mu}e_{J}^{\nu}\right)\), \(R_{\mu\nu}^{JK}(\omega) := \partial_{[\mu}\omega_{\nu]}^{JK} + \omega_{[\mu]}^{JK}\omega_{\nu]L}^{J}\) and \linebreak \(\tilde{R}_{\mu\nu}^{JK}(\omega) := \frac{1}{2} \epsilon_{JKLM}R_{\mu\nu LM}(\omega)\).

Here, with \(\gamma^{-1}\), the second term is the Holst term; while, with \(\gamma = -i\), this Lagrangian density gives the complex value \(SU(2)\) Ashtekar connection. For \(\gamma = 1\), one gets the real valued \(SU(2)\) Barbero connection [55]. This fact is already discussed in the first section.

The expression of the Nieh - Yan density is given as [55] 

\begin{eqnarray}
I_{\mathcal{NY}} = \epsilon^{\mu\nu\beta\eta} \left[D_{\mu}(\omega)e_{\nu}^{J} D_{\beta}(\omega) e_{J\beta} - \frac{1}{2}\Sigma_{\mu\nu}^{JK}R_{\beta\eta JK}(\omega)\right]
\end{eqnarray}   

Where, \( D_{\mu}(\omega)e_{\nu}^{J} = \partial_{\mu} e_{\nu}^{J} + \omega_{\mu K}^{J}e_{\nu}^{K}\). For a torsion-free connection, the
Nieh-Yan density disappears.

 In this proposal , in the time gauge, Nieh-Yan topological density of a theory of gravity permits to explain gravity as a real \(SU(2)\) connection. For \(\gamma = 1\), the set of constraints for both the Hamiltonian and the Barbero formalism are same. For the rest value of the BI parameter \(\gamma\), the Immirzi formulation exhibits \(\frac{1}{\gamma}\). This parameter \(\gamma\) is analogous to \(\theta\) parameter of the quantum chromodynamics. This parameter (\(\gamma\)) implies an enriched vacuum structure of gravity. The Nieh-Yan density completely constructed from  geometric quantities; while, the modified Holst terms exhibits fields of matter. With the aid of connection equation of motion both are connected [55].

\subsubsection{Pros and cons of Topological interpretation of the BI parameter} 

In this proposal, the authors gave the interpretation of the BI parameter \(\gamma\) topologically using the Nieh - Yan density with the real valued \(SU(2)\) connection. From this proposal, the BI parameter \(\gamma\) can be compared with the \(\theta\) parameter of the quantum chromodynamics.   

\subsection{The Peccei-Quinn mechanism in gravity and the nature of the BI parameter}

Promoting the BI parameter \(\gamma\) as a field was actively research area between 2007 to 2011; that added a new physical significance for the BI parameter \(\gamma\) with topological perspective in LQG. 

In 2009, Mercuri [56] gave a proposal on nature of the BI parameter \(\gamma\) using Peccei-Quinn mechanism in gravity in which the BI parameter \(\gamma\) is taken as a field. Using the Holst formalism, the modified Hilbert - Palatini action is  obtained. This modified action with the Nieh - Yan invariant (spacetime with the torsion) and the matter coupling is given as \(S_{Tot} = S_{HP}[e, \omega] +S_{NY}[e, \omega] + S_{mat} \). Hence, 

\begin{eqnarray}
\begin{aligned}
S_{Tot} = & \ - \frac{1}{16 \pi G} \int e_{c} \wedge e_{d} \wedge  \gamma R^{cd} + \frac{\gamma}{16 \pi G } \int \left(T^{c}\wedge T_{c} - e_{c} \wedge e_{d} \wedge R^{cd}\right) \\ & + \frac{i}{2} \int \star e_{c} \wedge \left(\bar{\psi} \gamma^{c}D\psi - \overline{D\psi}\gamma^{c}\psi + \frac{i}{2} m e^{c} \bar{\psi}\psi\right)
\end{aligned}
\end{eqnarray}

Here, the author suggested that to promote the BI parameter \(\gamma\) as a field; the contribution from divergence to the chiral anomaly must be reabsorbed [56].

The author also implied that the Peccei-Quinn mechanism (this mechanism is used for conservation of CP for strong force in which pseudo particle effect are considered with a scalar field) permits one to connect the constant value of the BI parameter \(\gamma\) to other certain topological ambiguities. This connection creates an interaction between gravity and the field of the BI parameter [56].

From the spontaneous symmetry breaking (\(SU(2) \times U(1)\)), the obtained quark mass matrices \(M\) is not diagonal and Hermitian. A chiral rotation is needed to diagonalize it. Similar to this, The chiral rotation of the fermionic measure in the Euclidean path-integral produces a NY term with Pontryagin class in spacetime with torsion that is diverged as the square
of the regulator [56].

\begin{eqnarray}
\delta \psi \delta \bar{\psi} \rightarrow  \delta \psi \delta \bar{\psi} exp \lbrace \frac{i}{8 \pi^{2}} \int \beta \left[R_{cd}\wedge R^{cd} + 2 \Lambda^{2} \left(T_{c}\wedge T^{c} - e_{c} \wedge e_{d} \wedge R^{cd}\right)\right]\rbrace
\end{eqnarray}

Where, \(\beta\) is the transformation parameter, \(\Lambda\) is a regulator.

\begin{eqnarray}
\begin{aligned}
S_{eff} = & \ S_{HP}[e, \omega] + S_{D} [e, \omega, \psi, \bar{\psi}] + \frac{1}{8 \pi^{2}} \beta \int R_{cd}\wedge R^{cd} + \frac{1}{16 \pi G} \left(\gamma + \frac{4G}{\pi} \beta \Lambda^{2}\right) \\ & \times \int \left(T^{c}\wedge T_{c} - e_{c} \wedge e_{d} \wedge R^{cd}\right)
\end{aligned}
\end{eqnarray}

Any attempt  of removal of the regulator results into the divergence of \(\beta\). By promoting the BI parameter \(\gamma\) as a field this divergence is reabsorbed [56]. Hence,

\begin{eqnarray}
\begin{aligned}
S_{eff} = & \ S_{HP}[e, \omega] + S_{D} [e, \omega, \psi, \bar{\psi}]  + \frac{1}{16 \pi G} \\ & \int  \gamma^{'}(x) \left(T^{c}\wedge T_{c} - e_{c} \wedge e_{d} \wedge R^{cd}\right)
\end{aligned}
\end{eqnarray}

Where, \(\gamma^{'}(x) =\gamma(x) + \frac{4G}{\pi} \beta \Lambda^{2} \).

\subsubsection{Pros and cons of the Peccei-Quinn mechanism in gravity and the nature of the BI parameter}

In this proposal, the BI parameter \(\gamma\) is promoted as field using the Peccei-Quinn mechanism in LQG. This notion is important, since, it removes divergence from the total effective action. However,  the origin of the complex valued the BI parameter \(\gamma\) is still open to explore. 

\subsection{The Kodama state and the BI parameter}

In year 2006, Randono [57-58] wrote two paper on generalization of Kodama states in which he used the real valued BI parameter \(\gamma\) to generalize these states and derived physical interprtation of these states. The Kodama state is special in providing an exact solution to all the normal constraints of canonical quantum gravity. 

The Kodama state has an unambiguous semi-classical interpretation as a quantum  sort of  a classical spacetime (anti de Sitter space). Though, the structure of the phase space is complex. Therefore,  a generalization of the real valued BI parameter \(\gamma\) state is needed [57].

The state of Lorentzian Kodama is a solution to the quantum constraints in the Ashtekar formalism in which  the connection is complex valued. However, to get the classical general relativity one has to execute the reality conditions ensuring the reality of the metric [57].

In the Euclidean framework
formalism, the \(SO(4)\) group  is divided into two left and right parts  as in the complex framework. The Ashtekar variables exhibits a real valued \(SO(3)\) connection and
its real valued momentum conjugate for the left handed part of the group . The corresponding state in the Euclidean framework is a pure phase; because, the connection is real. Thus, the state is written as [57], 

\begin{eqnarray}
\Psi[A] = \mathcal{N}e^{-\frac{3}{4\kappa\lambda}} \int Y_{\mathcal{CS}}[A]
\end{eqnarray}

Where, \( \int Y_{\mathcal{CS}} [A]\) is the Chern - Simon term, 

For this result, no reality condition is required; since, the structure of the  phase space of the Euclidean framework is simple. Hence, the complexification of the phase space (in which the BI parameter \(\gamma\) is the complex valued \(\gamma = i\)) is the main cause for the requirement of the reality conditions [57].

The Kodama state begins with the Holst action with the cosmological constant is given as [57], 

\begin{eqnarray}
S_{H} = \frac{1}{\kappa} \int_{\mathcal{M}} \star e \wedge e \wedge R + \frac{1}{\gamma} e \wedge e \wedge R - \frac{\lambda}{3} \star e \wedge e \wedge e \wedge e 
\end{eqnarray} 

Where, \(\gamma\) is the BI parameter,  \(e = \frac{1}{2} \gamma_{J}e^{J}\), \(R = \frac{1}{4} \gamma_{[J}\gamma_{K]}\omega^{JK}\), \(\star = -i \gamma^{5} = \gamma^{0}\gamma^{1}\gamma^{2}\gamma^{3}\).

The Chiral symmetric Holst action written as [57], 

\begin{center}
\(S = \frac{1}{\kappa} \int_{\mathcal{M}} 2 \left(\alpha_{L}P_{L} + \alpha_{R}P_{R}\right) \star \Sigma \wedge \left(R - \frac{\lambda}{6}\right)\) 
\end{center}

\begin{center}
= \(\frac{2}{\kappa} \int_{\mathcal{M}} \alpha_{L}\star \Sigma_{L}\wedge \left(R_{L} - \frac{\lambda}{6}\Sigma_{L}\right) + \alpha_{R}\star \Sigma_{R}\wedge \left(R_{R} - \frac{\lambda}{6}\Sigma_{R}\right)\)
\end{center}

If \(\alpha_{L} + \alpha_{R} = 1\) and \(\gamma = \frac{-i}{\alpha_{L} - \alpha_{R}} \); then, the equation becomes [57], 

\begin{eqnarray}
S = \frac{1}{\kappa} \int_{\mathcal{M}} \left(\alpha_{L} + \alpha_{R}\right) \star \Sigma \wedge \left(R - \frac{\lambda}{6} \Sigma\right) + i\left(\alpha_{L} - \alpha_{R}\right) \Sigma \wedge R
\end{eqnarray}

In year 2011, Wieland [59] gave a proposal namely, - Complex Ashtekar variables, the Kodama state and spinfoam gravity; in which he used the complex valued Ashtekar variable and the real valued BI parameter \(\gamma\). He used  \(SL(2,\mathbb{C})\) Kodama state and proposed a spinfoam vertex amplitude.

As per usual method, it also start with the Holst action with the cosmological constant [59]; i.e., 

\begin{eqnarray}
S [e, \omega] = \frac{\hbar}{4 l_{P}^{2}} \int_{\mathcal{M}} e^{J} \wedge e^{K} \wedge \left(\epsilon_{JKLM}R^{LM}[\omega] - \frac{2}{\gamma}R_{JK}[\omega]-\frac{\Lambda}{6}\epsilon_{JKLM} e^{L}\wedge e^{M}\right)
\end{eqnarray}

Where, \(\gamma\) is the BI parameter (\(\gamma \in \mathbb{R}\)).

As mentioned, in this proposal, the real valued BI parameter \(\gamma\) and the complex valued Ashtekar variable is considered [59].

\subsubsection{Pros and cons of the Kodama state and the BI parameter}

The generalization of the Kodama state can only be done with the real valued BI parameter \(\gamma\); since, the BI parameter \(\gamma\) with the complex value makes the state complex and ambiguous. Here, the significance of the complex valued BI parameter is also not clear.

\subsection{The quantum gravity BI parameter - a general physical and topological interpretation}

In year 2013, El Naschie [60] gave a proposal on general physical and topological interpretation of the BI parameter \(\gamma\). This proposal is not directly related to LQG. In this paper, he explained the BI parameter \(\gamma\) of LQG as a definite quantum entanglement correction. 

According to this proposal, The BI parameter \(\gamma\) is not only
a free basic parameter of LQG; but also, it is an exact sort of a basic constant of the micro-spacetime topology [60].

As mentioned, in the section (2.3), one of the fixed value of the BI parameter \(\gamma\) from black hole entropy calculation in LQG is given as [60]

\begin{eqnarray}
\gamma = \frac{log 2}{\pi \sqrt{3}} = 0.055322
\end{eqnarray}

From Hardy's quantum entanglement, the author proposed that the value of the BI parameter \(\gamma\) is same as that is obtained from  quantum entanglement correction [60]. i.e., 

\begin{eqnarray}
\gamma = \phi^{6} = \left(\frac{\sqrt{5} - 1}{2}\right)^{6} = 0.055728
\end{eqnarray}

\subsubsection{Pros and cons of the quantum gravity BI parameter — a general physical and topological interpretation}

This proposal advocates the real valued BI parameter \(\gamma\). It does not explain the physical significance of the BI parameter \(\gamma\) in LQG. 

\subsection{A correction to  the BI parameter of SU(2) spin networks}

In year 2014, Sadiq [61] gave a correction to the BI parameter \(\gamma\) of \(SU(2)\) spin networks. In this paper, by  taking \(j=1\) and to preserve   \(SU(2)\) symmetry of theory; twice value of the BI parameter \(\gamma\) is proposed.  Previously in LQG, the BI parameter \(\gamma\) is fixed by \(j=1\) transitions of spin networks as the dominant processes instead of \(j = 1/2\) transitions. It means \(SO(3)\) should be a gauge group instead of \(SU(2)\).

This proposal begins with [35] (see section 2.3) and gave a correction to the BI parameter \(\gamma\). The author investigated that if \(SU(2)\) is the compatible gauge group and \(j_{min} =1\) process governs; then the change in mass of the black hole during the transition is [61] 

\begin{eqnarray}
\Delta M = 2 \hbar \omega_{QNM} 
\end{eqnarray}

Since, the value of \(\omega_{QNM}\) is \(\omega_{QNM} = \frac{ln 3}{8 \pi M}\); the change in mass is [61], 

\begin{eqnarray}
\Delta M = \frac{ 2 \hbar ln 3}{8 \pi M}
\end{eqnarray}

Therefore, the fixed value of \(\gamma\) for \(j_{min} = 1\) is  [61]

\begin{eqnarray}
\gamma = \frac{ln 3}{\pi \sqrt{2}}
\end{eqnarray}

\subsubsection{Pros and cons of a correction to  the BI parameter of SU(2) spin networks} 

In this proposal, the author modified  the fixed value of BI parameter \(\gamma\) that was proposed in [35]. this proposal advocates the real valued BI parameter \(\gamma\). The physical significance is based on the black hole entropy calculation in LQG.

\subsection{Physical effect of the Immirzi parameter in loop quantum gravity}

In this proposal, Perez and Rovelli [62] proposed that the BI term in the action (Holst) does not disappear on the shell when fermions are there. The BI term is also present in the equations of motion. The BI parameter \(\gamma\) governs the coupling constant of a four-fermion interaction (it is mediated by a torsion). In other words, The BI parameter \(\gamma\) is a coupling constant that governs the strength of a four-fermion interaction. Thus the BI parameter \(\gamma\) may show physical effects that can be observed independently from LQG.

The Holst action with the fermionic field is expressed as [62], 

\begin{eqnarray}
S[e, A, \psi] = S[e, A] + \frac{i}{2} \int d^{4} x e \left(\bar{\psi} \gamma^{J} e_{J}^{c}D_{c}\psi - \overline{D_{c} \psi} \gamma^{J}e_{J}^{c} \psi\right)
\end{eqnarray}

Where, \(S[e, A] = \frac{1}{16 \pi G}\left(\int d^{4} x e e_{J}^{c}e_{K}^{d}F_{cd}^{JK} - \frac{1}{\gamma} \int d^{4} x e e_{J}^{c}e_{K}^{d} * F_{cd}^{JK}\right)\), \(D_{c}\) is a covariant derivative and \(\gamma^{J}\) is the Dirac matrices. In this proposal, \(D_{[c}e_{d]}^{J}\) as the fermionic current. In the connection, the fermion current behaves as a source for a torsion component [62]. 

In the fermionic current, the linear term  are total derivative,; hence, the resulting action is, 

\begin{center}
\(S[e,\psi] = S[e] + S_{f}[e, \psi]  + S_{int}[e,\psi]\)
\end{center}

\begin{eqnarray}
\begin{aligned}
\therefore S[e,\psi]  = & \ \frac{1}{16 \pi G} \int d^{4} x e e_{J}^{c}e_{K}^{d}F_{cd}^{JK} [\omega [e]] + i \int d^{4} x e \bar{\psi} \gamma^{J}e_{J}^{c} D_{c}[\omega [e]]\psi \\ & - \frac{3}{2} \pi G \frac{\gamma^{2}}{\gamma^{2}+1}\int d^{4} x e \left(\bar{\psi} \gamma_{5}\gamma_{A} \psi\right)\left(\bar{\psi} \gamma_{5}\gamma^{A} \psi\right)
\end{aligned}
\end{eqnarray}

the standard coupling of the Einstein-Cartan theory is recovered in third term with the limit \(\gamma \rightarrow \infty\) [62].

\subsubsection{Pros and cons of physical effect of the Immirzi parameter in loop quantum gravity}

In this proposal, The BI parameter \(\gamma\) is a coupling constant that governs the strength of a four-fermion interaction. Here, The BI parameter \(\gamma\) is indeed free parameter. \(\gamma = i\) gives the self-dual Ashtekar canonical formalism and \(\gamma = 1\) (real valued) gives the \(SU(2)\) Barbero connection.

\subsection{A relation between the Barbero-Immirzi parameter and the standard model}

In year 2010, Broda and Szanecki [63] established the relationship between the BI parameter \(\gamma\) and the standard model.

In this proposal, the authors used  Sakharov’s method with the NY (Nieh - Yan) term (in the Holst action) that fixed the BI parameter \(\gamma\) by considering the Lagrangian density [63], i.e, 

\begin{eqnarray}
\mathcal{L} = \alpha \star \left(e^{c} \wedge e^{d}\right) \wedge R_{cd} - \beta \left(T^{c}\wedge T_{d} - e^{c} \wedge e^{d} \wedge R_{cd}\right)
\end{eqnarray}

Where, \(\gamma = \frac{\alpha}{\beta}\)

The Einstein - Hilbert Lagrangian is written as [63], 

\begin{eqnarray}
\mathcal{L}_{EH} = - \frac{1}{12} \left(\frac{M}{4 \pi}\right)^{2} \left(N_{0} + N_{\frac{1}{2}} - 4 N_{1}\right) \star \left(e^{c}\wedge e^{d}\right) \wedge R_{cd}
\end{eqnarray} 

Where \(N_{1}\) is the gauge fields  number, \(N_{\frac{1}{2}}\) two-component fermion fields number, and \(N_{0}\) is the minimal scalar degrees of freedom number. Here the BI parameter \(\gamma\) is defined as  [63]

\begin{eqnarray}
\gamma = \frac{- \frac{1}{12}\left(N_{0} + N_{\frac{1}{2}} - 4 N_{1}\right)}{-\frac{1}{4}N_{L}} = \frac{1}{9} \approx 0.11
\end{eqnarray}

Here, \(N_{L}\) is chiral left-handed modes number. By taking \(N_{0} = 4 \) (for Higgs),   \(N_{1} = 12\),   \(N_{\frac{1}{2}} = 45\) and \(N_{L} = 3\) (3 neutrinos) [63].

which is approximately equal to (one of the value of the BI parameter \(\gamma\) in black hole entropy calculation in LQG)[63].

\begin{eqnarray}
\gamma = \frac{\ln 2}{\pi \sqrt{3}}  \approx 0.13
\end{eqnarray}

\subsubsection{Pros and cons of a relation between the Barbero-Immirzi parameter and the standard model}

This proposal established a relation between the BI parameter \(\gamma\) and the standard model. it advocates the real valued BI parameter \(\gamma\) by comparing the results with the black hole entropy calculation in LQG.  

\subsection{The holographic principle and the BI parameter of loop quantum gravity}

In year 2015,  Sadiq [64] gave a proposal that correlates the BI parameter \(\gamma\) in LQG and the holographic principle. In this proposal the BI parameter \(\gamma\) is fixed using the equipartition theorem based on LQG at holographic boundary in such a way that Unruh - Hawking law of temperature holds and follows. such derived value of the BI parameter \(\gamma\) exhibits validity universely. In that way this approach correlates the value of the BI parameter \(\gamma\) in LQG and the holographic principle. In this proposal, the real valued BI parameter is considered. Since, the relation between the holographic principle and LQG demands more research; this proposal is very briefly given.

\subsubsection{Pros and cons of the holographic principle and the BI parameter of loop quantum gravity}

Similar to majority proposals, this proposal also advocates the real valued BI parameter.

\subsection{Discussion}  

The BI parameter \(\gamma\) is a free parameter as well as an enigmatic parameter of LQG. It is a free parameter; because, it can be real valued as well as complex valued. It is enigmatic parameter; because, its significance with either real value or with complex value, is still obscure.

As mentioned, there are two sort of version for the connection variables: \(SL(2,\mathbb{C})\) with a self duality of Ashtekar formalism \(\gamma = i\) and the other is the connection with a real \(SU(2)\) Barbero connection \(\gamma = 1\). 

If LQG is compared with the other quantum gravity theory; it has only one free parameter in its formalism. Here, several proposals on the physical significance of the BI parameter \(\gamma\) is discussed. Some argues in favor of the real valued BI parameter; while, the others argues in favor of the complex valued BI parameter. The real valued proposals on BI parameter are more tangible. The value of the BI parameter found from the Black hole entropy calculation in LQG, has more consent then other proposals; because, the same value is applied for all sort of black holes. 

Still, the exact physical significance of the BI parameter \(\gamma\) is yet to be found. There are several questions regarding the choice of the BI parameter are discussed via various proposals. Most importance question is regarding the physical significance of the area operator and the volume operator spectrum with the complex valued BI parameter. The complex valued BI parameter is also important; because, it removes the mathematical complexities from the equations of the constraints, especially from the Hamiltonian constraint. Research on the physical significance of the complex valued BI parameter \(\gamma = i\) will open up a new direction in the field of the quantum gravity. Time will shed light on these mysteries.  
 
\section{Concluding remarks}

\begin{itemize}

 \item In this paper, initially, a short introduction of the Barbero - Immirzi parameter \(\gamma\) along with the introduction to Ashtekar formalism, origin of the BI parameter, the Holst action and historical timeline of research in the physical significance of the BI parameter \(\gamma\) in LQG are given.

 \item Since, the value of the BI parameter \(\gamma\) and its implication is very important, especially in the area operator spectrum and black hole entropy calculation in LQG; afterwards, these are elaborated. 
 
 \item At last, various proposal on the physical significance of the BI parameter \(\gamma\) in LQG are briefly given with their pros and cons.
 
 \item Hence, the BI parameter \(\gamma\), whether complex-valued or real-valued is a crucial free parameter of LQG. 
 
\end{itemize}

\section{Acknowledgement}

The authors are thankful to Physics Department, Saurashtra University, Rajkot.

\end{document}